\documentclass[prl,aps,twocolumn,superscriptaddress,showpacs]{revtex4}
\usepackage{epsfig}
\usepackage[draft]{hyperref}

\begin{document}
\title{Stirring Potential for Indirect Excitons}

\author{M.~W.~Hasling}
\affiliation{
Department of Physics, University of California at San Diego, La Jolla, CA 92093-0319
}
\author{Y.~Y.~Kuznetsova}
\affiliation{
Department of Physics, University of California at San Diego, La Jolla, CA 92093-0319
}
\author{P.~Andreakou}
\affiliation{
Department of Physics, University of California at San Diego, La Jolla, CA 92093-0319
}
\affiliation{
Laboratoire Charles Coulomb, Universit{\'e} Montpellier 2, CNRS, UMR 5221, F-34095 Montpellier, France
}
\author{J.~R.~Leonard}
\affiliation{
Department of Physics, University of California at San Diego, La Jolla, CA 92093-0319
}
\author{E.~V.~Calman}
\affiliation{
Department of Physics, University of California at San Diego, La Jolla, CA 92093-0319
}
\author{C.~Dorow}
\affiliation{
Department of Physics, University of California at San Diego, La Jolla, CA 92093-0319
}
\author{L.~V.~Butov}
\affiliation{
Department of Physics, University of California at San Diego, La Jolla, CA 92093-0319
}
\author{M.~Hanson}
\affiliation{
Materials Department, University of California at Santa Barbara, Santa Barbara, CA 93106-5050
}
\author{A.~C.~Gossard}
\affiliation{
Materials Department, University of California at Santa Barbara, Santa Barbara, CA 93106-5050
}

\begin{abstract}
We demonstrate experimental proof of principle for a stirring potential for indirect excitons. The azimuthal wavelength of this stirring potential is set by the electrode periodicity, the amplitude is controlled by the applied AC voltage, and the angular velocity is controlled by the AC frequency.
\end{abstract}
\date{\today}
\maketitle

Controlled moving potentials are used for studying transport properties of excitons. An established method for creating moving potentials for excitons is based on surface acoustic waves (SAW). Transport of excitons, exciton-polaritons, and laterally separated electrons and holes via SAW has been realized~\cite{Rocke97, Rudolph07, Lazic10, Cerda-Mendez10, Violante14, Lazic14}. Large exciton transport distances in moving potentials can be achieved with indirect excitons composed of electrons and holes in spatially separated layers [Fig.~1(a)]: Indirect excitons have long lifetimes and can travel over large distances before recombination~\cite{Hagn95, Butov98, Larionov00, Gartner06, Ivanov06}. Effective transport of indirect excitons via SAW was demonstrated recently~\cite{Rudolph07, Lazic10, Violante14, Lazic14}.

Moving potentials for indirect excitons can also be created by laterally modulated AC voltage patterns. Indirect excitons have a built-in dipole moment $ed$ ($d$ is the separation between the electron and hole layers) so voltage-controlled electric field perpendicular to the QW plane $F_z$ creates the desired potential landscape for indirect excitons $E(x,y) = - e d F_{\rm z}(x,y)$~\cite{Miller85}. Transport of indirect excitons was studied in static potentials formed by time-independent laterally modulated $F_z$, including ramps~\cite{Hagn95, Gartner06, Leonard12}, lattices~\cite{Remeika09}, traps~\cite{High09prl, Vogele09}, and circuit devices~\cite{Andreakou14}. Transport of indirect excitons in a moving electrostatic lattice potential -- conveyer -- was recently demonstrated~\cite{Winbow11}. The conveyer was created by applying AC voltages to the electrodes of an electrostatic lattice potential for excitons. This gave a traveling lattice moving indirect excitons laterally across the sample. The excitonic conveyer moves indirect excitons as charged coupled devices move electrons~\cite{Smith10}. The moving wave of voltage couples to the dipole moment of indirect excitons in the former~\cite{Winbow11} and charge of electrons in the latter~\cite{Smith10}.

\begin{figure}[h]
\centering
\includegraphics[width=3in]{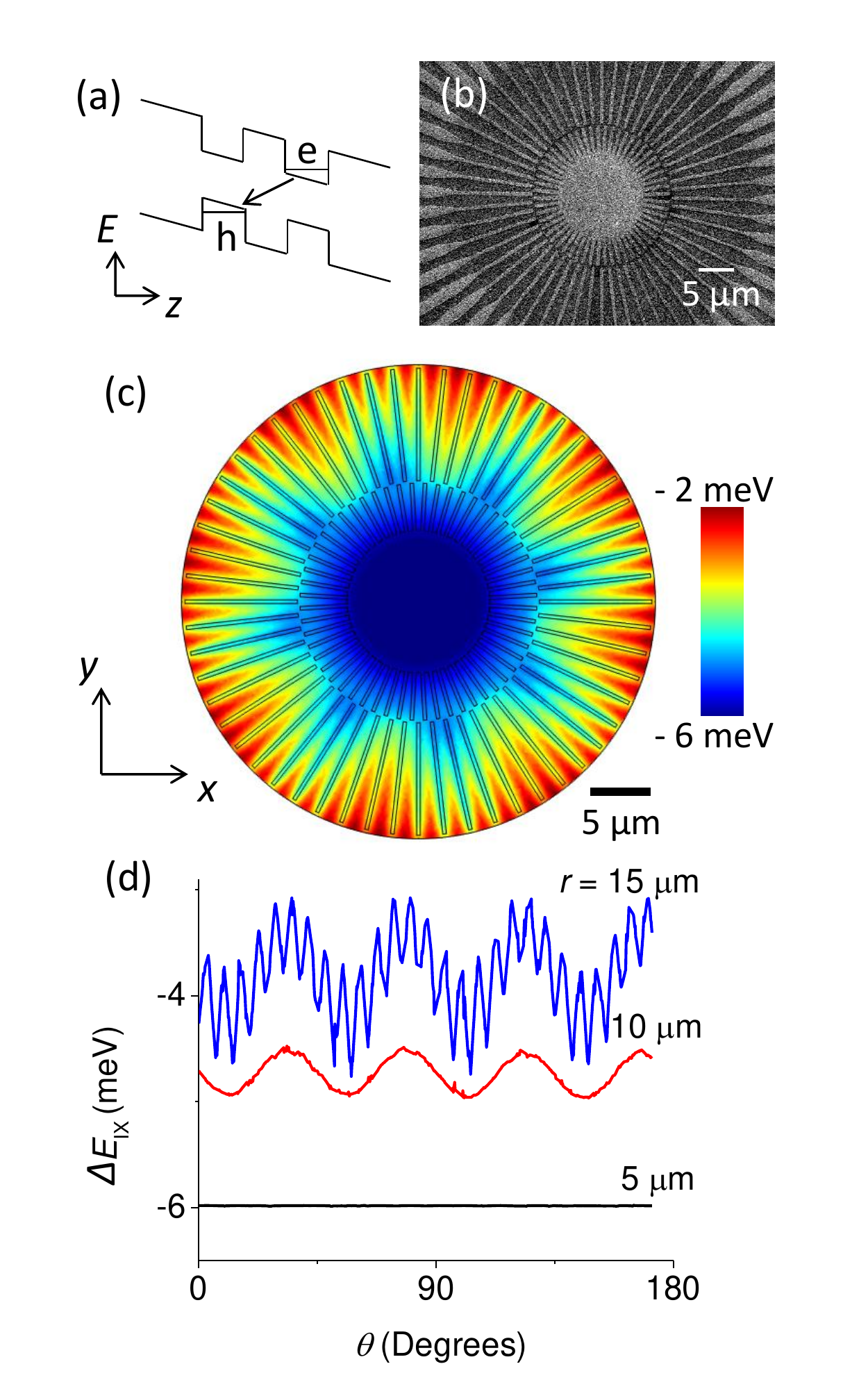}
\label{fig1}
\caption{(a) CQW diagram. e, electron; h, hole. An indirect exciton is shown by an arrow. (b) SEM image of the electrode pattern forming the stirring potential for indirect excitons. (c) Snapshot of calculated indirect exciton energy shift $E(x,y)=- e d F_{\rm z}$ in the stirring potential. (d) Angular cuts of calculated snapshot of indirect exciton energy shift at different radii. DC voltage on center and outer electrodes $V_{\rm c} = V_{\rm o} = .5$~V, AC voltage on outer electrodes $V_{\rm AC} = 60$~mV. The stirring potential for indirect excitons rotates with the angular velocity $\omega_{\rm car} = f \theta_{\rm car}$ where $f$ is the AC frequency and $\theta_{\rm car}$ is the angular wavelength of the potential.}
\end{figure}

Voltage control allows the realization of a variety of moving potentials for indirect excitons. Stirring potentials are of particular interest. They can be used for studying rotating exciton matter. Optical and exciton-polariton systems with optically generated or spontaneously formed vorticity were intensively investigated, see \cite{Scheuer99, Padgett04, Lagoudakis09, Krizhanovskii10, Roumpos11} and references therein. Stirring was also explored in the studies of rotating atom matter, see \cite{Abo-Shaeer01, Schweikhard04, Hadzibabic06} and references therein. In these diverse systems, stirring was explored to generate and study vortical states with angular momentum.

In this work, we present experimental proof of principle for a stirring potential for indirect excitons. The indirect excitons are created in a GaAs coupled quantum well structure (CQW) grown by molecular beam epitaxy [Fig. 1(a)]. An $n^+$-GaAs layer with $n_{\rm Si} = 10^{18}$~cm$^{-3}$ serves as a homogeneous ground plane. Two 8 nm GaAs QWs are separated by a 4 nm Al$_{.33}$Ga$_{.67}$As barrier and positioned 100~nm above the $n^+$-GaAs layer within an undoped 1~$\mu$m thick Al$_{.33}$Ga$_{.67}$As layer. Positioning the CQW closer to the homogeneous electrode suppresses the in-plane electric field \cite{Hammack06jap}, which otherwise can lead to exciton dissociation. Time-dependent voltage is applied to an electrode pattern deposited on top of the sample, resulting in a variable $F_z$ in the CQW plane, and, in turn, a controllable potential landscape for indirect excitons $E(x,y,t)$.

A stirring (carousel) potential is created by a centrally symmetric set of semitransparent 11~nm thick Ti--Pt--Au electrodes on the sample surface. The electrode width is 300 nm, carousel periodicity is 7 electrodes, and the angular wavelength of the carousel potential is $\theta_{\rm car} = 45$ degrees [Fig.~1(c)]. The carousel electrodes are covered by a layer of transparent insulation (300~$\mu$m thick SiO$_2$). A set of connecting electrodes (2~$\mu$m wide, 200~nm thick ITO) provides the contacts to the carousel electrodes through $3 \times 3$~$\mu$m openings in the insulating layer that allows the construction of the periodically connected set of carousel electrodes \cite{SI}.

AC voltages to the carousel electrodes are applied by coaxial cables with impedance-matching termination at the sample \cite{SI}. The cable bandwidth complies with the frequency used in the experiments. The regime where the indirect excitons form the ground state, i.e. have lower energy than spatially direct excitons in the CQW, is realized by DC biases $V_{\rm c}$ and $V_{\rm o}$ applied to the central electrode and outer set of electrodes, respectively. For $V_{\rm c} = V_{\rm o}$ used in the experiments, decreasing electrode density toward the carousel edges reduces $F_{\rm z}$ and, in turn, exciton energy shift, producing a confining potential for indirect excitons with the exciton energy reducing toward the carousel center [Figs.~1(c), 1(d)]~\cite{Kuznetsova10}. A set of differentially phase-delayed AC voltage sine waves applied to the outer electrodes at frequency $f = 47.5$~MHz creates a stirring potential for indirect excitons -- the excitonic carousel rotating with the angular velocity $\omega_{\rm car} = f \theta_{\rm car} \sim 2 \times 10^9$ degrees per second. The amplitude of the carousel potential for indirect excitons is controlled by the AC voltage $V_{\rm AC}$. The radial dependence of the snapshot of the carousel potential is shown in Fig.~1(d). An effective stirring potential for indirect excitons is realized at the distance $r \sim 10$~$\mu$m from the center. At large $r \gtrsim 15$~$\mu$m, the sinusoidal envelope of the carousel potential with $\theta_{\rm car} = 45$ degrees is modulated by $\sim 6$ degrees-period ripples, which originate from the finite spacing between the carousel electrodes $d_{\rm s}$. The amplitude of these ripples is smaller for a smaller $d_{\rm s}$, and the ripples essentially vanish at $r \sim 10$~$\mu$m where $d_{\rm s} \lesssim 0.5$~$\mu$m for the structure [Figs.~1(c), 1(d)].

The sample mounts in a He cryostat at 1.5 K. The excitons are photoexcited by 700~nm Ti:sapphire laser focused to a spot $\sim 6$~$\mu$m in diameter at $r \sim 10$~$\mu$m. The exciton density is controlled by the laser excitation power $P$. Photoluminescence (PL) images of the exciton cloud are taken by a CCD with a bandpass filter $800 \pm 5$~nm covering the spectral range of the indirect excitons. The diffraction-limited spatial resolution is 1.4~$\mu$m.

\begin{figure}[h]
\centering
\includegraphics[width=3in]{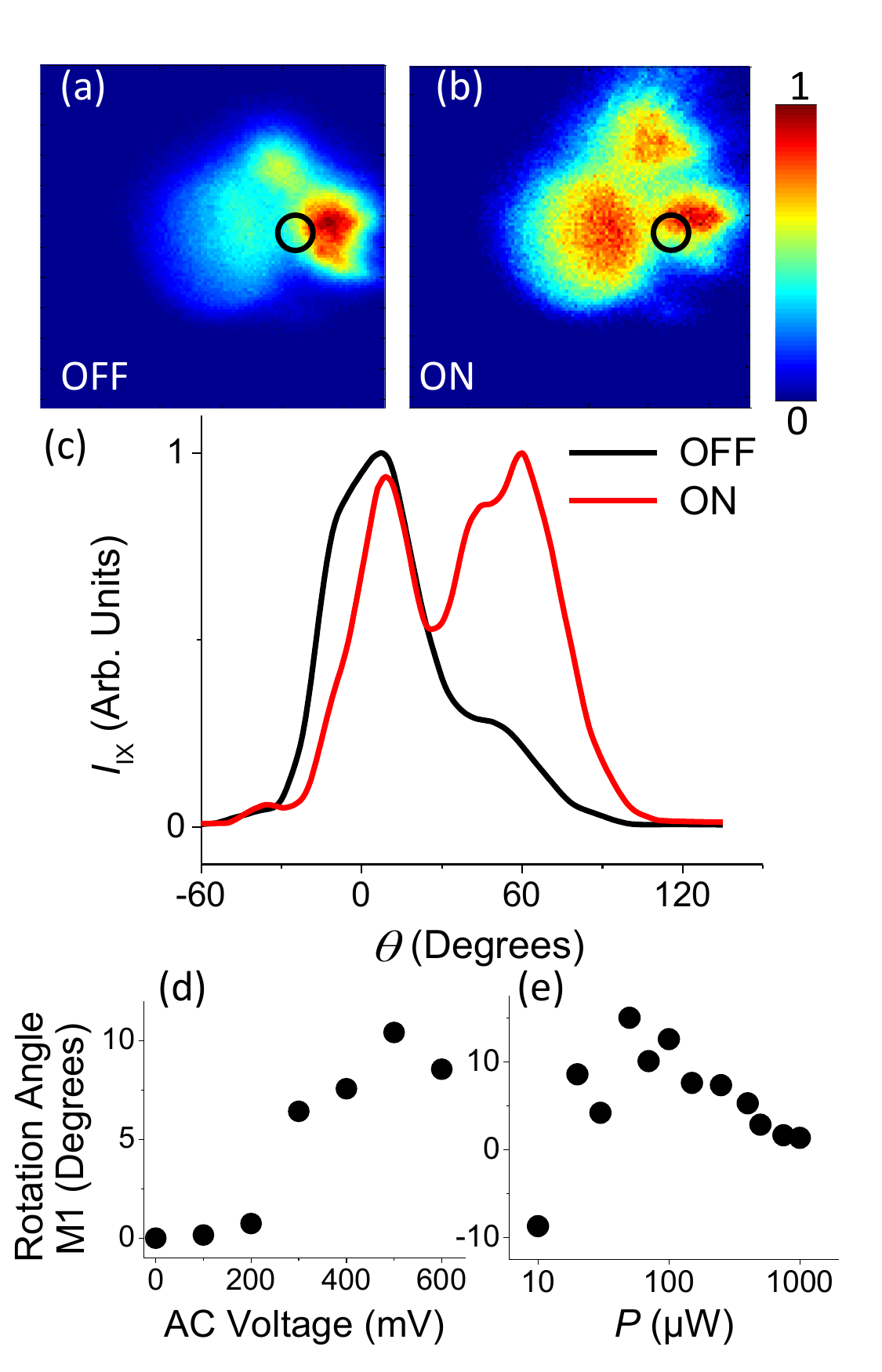}
\label{fig2}
\caption{(a)-(b) $x-y$ emission images of indirect excitons with AC voltage off and on. Excitation power $P = 50$~$\mu$W. Emission intensity in (b) is multiplied by a factor of 3.4 for clarity. Black circles show the excitation spot. The circle size corresponds to the excitation spot size. (c) Angular cuts of emission images in (a) and (b) at $r = 15$~$\mu$m. (d)-(e) The average rotation angle of indirect excitons via stirring potential M1 as a function of AC voltage $V_{\rm AC}$ (d) and excitation power $P$ (e). $V_{\rm AC} = 600$~mV in (a)-(c),(e). $P = 100$~$\mu$W in (d). For all data, DC center and outer voltage $V_{\rm c} = V_{\rm o} = 5$~V and AC frequency $f = 47.5$~MHz.}
\end{figure}

Figure~2 presents experimental proof of principle for a stirring potential for indirect excitons. Figures 2(a) and 2(b) show $x-y$ PL images of indirect excitons for carousel off and on, respectively. The corresponding PL intensity profiles $I(\phi)$ presenting angular cuts of the emission images at $r = 15$~$\mu$m are shown on Fig.~2(c). The rotation of the exciton cloud by the carousel potential is presented by the extension of the exciton cloud along the angle of the stirring potential. We quantify it by the first moment of the PL intensity ${\rm M_1} = \int \phi I(\phi)d\phi / \int I(\phi)d\phi$, which characterizes the average rotation angle of indirect excitons via carousel.

Figure~2(d) presents exciton cloud rotation via carousel as a function of the AC carousel voltage $V_{\rm AC}$ which controls
the amplitude of the stirring potential. For a shallow carousel, the exciton cloud extension M1 is not affected by the carousel rotation showing that the excitons do not follow the stirring potential. In contrast, at higher amplitude of the stirring potential, excitons are moved by the stirring potential. The exciton cloud starts to follow the conveyer and M1 changes from constant to increasing with $V_{\rm AC}$ at $V_{\rm AC} \sim 200$~mV [Fig.~2(d)]. These results can be understood as follows: When the carousel amplitude is smaller than the exciton interaction energy or amplitude of disorder (given by the intrinsic disorder in the structure and ripples in the conveyer potential), excitons are not localized in the minima of the rotating carousel potential, and therefore are not stirred by the carousel. When the carousel amplitude becomes larger than both the exciton interaction energy and disorder amplitude, excitons can localize in the minima of the rotating carousel potential. This results in efficient stirring of excitons via carousel. The transition between these two regimes is similar to the dynamical localization-delocalization transition for excitons in conveyers~\cite{Winbow11} and to the localization-delocalization transition for excitons in static lattices~\cite{Remeika09}: At low (high) amplitude of the stirring potential, excitons are dynamically delocalized (localized) in the stirring potential.

Figure~2(e) presents the dependence of exciton stirring via carousel on excitation density $P$. Efficient exciton stirring via carousel is achieved at intermediate densities and becomes less efficient at low and high densities. These results can be understood as follows: At low densities, the indirect excitons are localized in local minima of the disorder potential and hardly follow the rotating carousel. At the intermediate densities, excitons screen the disorder and can be efficiently stirred by the carousel. At the high densities, excitons screen the carousel potential that makes exciton stirring via carousel less efficient. In other words, excitons can efficiently follow the stirring carousel potential when the exciton drift angular velocity in the carousel is higher than the carousel angular velocity $\omega_{\rm drift} \gtrsim \omega_{\rm car}$. For $\omega_{\rm drift} \propto \mu U$ ($\mu$ is exciton mobility, $U$ is the amplitude of stirring potential), an efficient exciton stirring is realized when both $\mu$ and $U$ are high. Screening of disorder results in the enhancement of exciton mobility while screening of the stirring potential results in the reduction of the amplitude of the stirring potential.

For an outlook of this work, we note that stirring potentials for indirect excitons can be used to generate vortices in indirect excitons. The performance of stirring potentials is limited by the quality of the top electrode pattern that can be improved by increasing the pattern dimensions. This can be achieved in CQW structures where the distances between the CQW and top and bottom electrodes are increased~\cite{Kuznetsova10}. Increasing the lateral dimensions for the top electrode will make possible improving the quality of the lithography processing and smoothing out the ripples in the carousel potential thus improving the quality of the carousel potential profile. Furthermore, if required, the lifetime of indirect excitons in the device can be increased by increasing the separation between the QW layers. This forms the subject for future work.

In summary, we report on the realization of an electrostatic stirring potential for indirect excitons with controllable amplitude, angular wavelength, and angular velocity.

We thank A.T. Hammack, M. Remeika, M. Vladimirova, and A.G. Winbow for discussions. This work was supported by the U.S. DOE Office of Basic Energy Sciences under Award No. DE-FG02-07ER46449. P.A. was supported by EU ITN INDEX.

%\newpage

%\bibliographystyle{apsrev}
%Labels are numeric, entries are in order of citation.

\end{document}

\renewcommand{\thefigure}{S\arabic{figure}}

\title{Stirring Potential for Indirect Excitons

Supplementary materials}

\author{M.~W.~Hasling}
\affiliation{
Department of Physics, University of California at San Diego, La Jolla, CA 92093-0319
}
\author{Y.~Y.~Kuznetsova}
\affiliation{
Department of Physics, University of California at San Diego, La Jolla, CA 92093-0319
}
\author{P.~Andreakou}
\affiliation{
Department of Physics, University of California at San Diego, La Jolla, CA 92093-0319
}
\affiliation{
Laboratoire Charles Coulomb, Universit{\'e} Montpellier 2, CNRS, UMR 5221, F-34095 Montpellier, France
}
\author{J.~R.~Leonard}
\affiliation{
Department of Physics, University of California at San Diego, La Jolla, CA 92093-0319
}
\author{E.~V.~Calman}
\affiliation{
Department of Physics, University of California at San Diego, La Jolla, CA 92093-0319
}
\author{C.~Dorow}
\affiliation{
Department of Physics, University of California at San Diego, La Jolla, CA 92093-0319
}
\author{L.~V.~Butov}
\affiliation{
Department of Physics, University of California at San Diego, La Jolla, CA 92093-0319
}
\author{M.~Hanson}
\affiliation{
Materials Department, University of California at Santa Barbara, Santa Barbara, CA 93106-5050
}
\author{A.~C.~Gossard}
\affiliation{
Materials Department, University of California at Santa Barbara, Santa Barbara, CA 93106-5050
}

\maketitle

\subsection{Electrode pattern creating the carousel for indirect excitons}

\begin{figure}[h]
\centering
\includegraphics[width=7in]{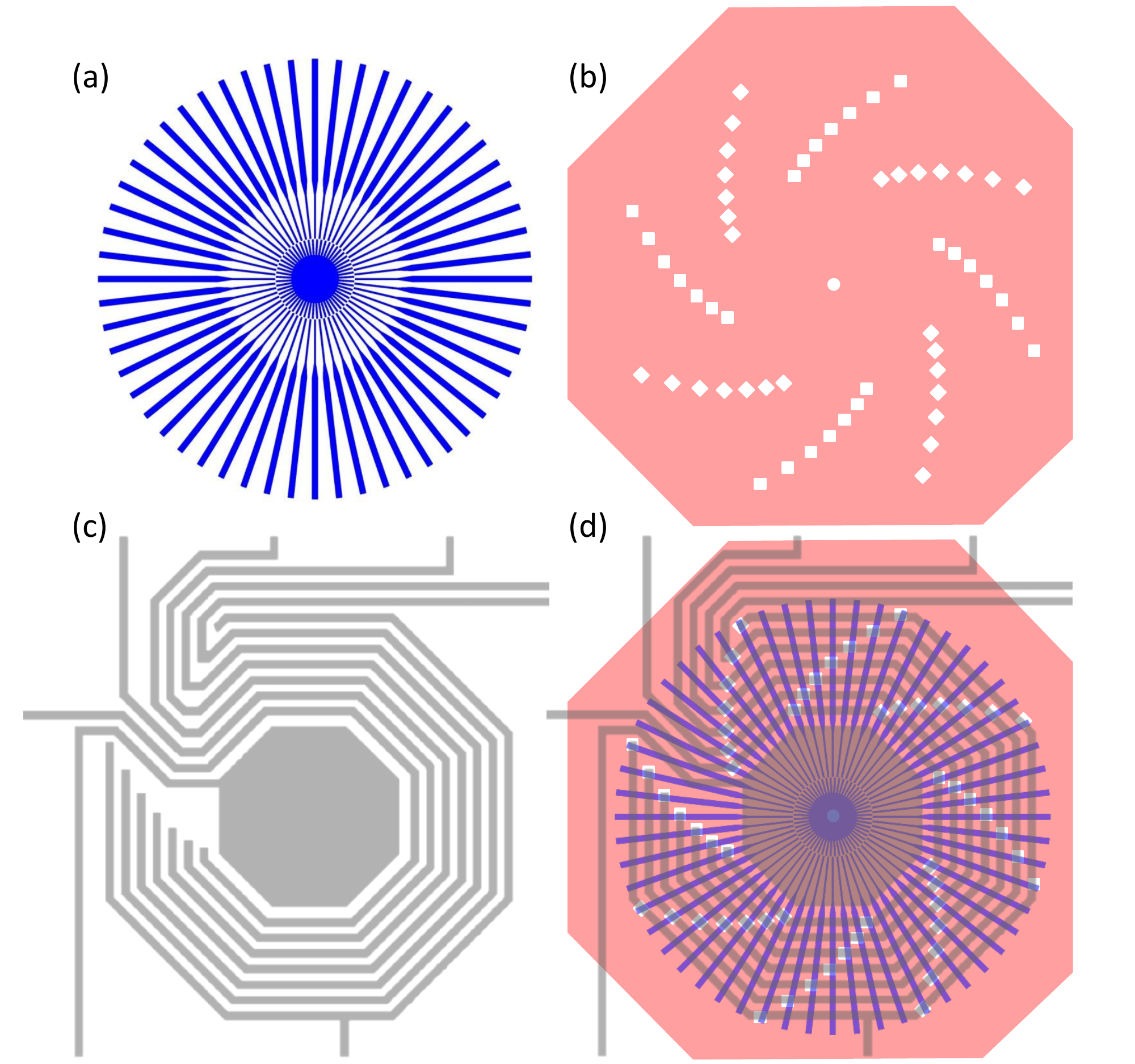}
\caption{Layers that form the stirring potential for indirect excitons. (a) Ti--Pt--Au carousel electrode layer. (b) SiO$_{2}$ insulating layer. (c) ITO connecting electrode layer. (d) The three layers overlaid. The holes in the insulating layer provide the contacts between the connecting electrodes and carousel electrodes.}
\end{figure}

The top electrodes are fabricated as follows. Mesas are etched to the conducting $n^+$-GaAs layer in order to connect to this homogeneous bottom electrode. Then the stirring potential is created with the carousel electrode layer, insulating layer, and connecting electrode layer (Fig.~S1). The carousel electrode layer [Fig.~S1(a)] is created by depositing 2~nm Ti -- 7~nm Pt -- 2~nm Au on the sample surface. This layer consists of a central electrode and separate 300~nm wide lines that form outer electrodes [Fig.~S1(a),~S2]. Static voltages $V_{\rm c}$ and $V_{\rm o}$ are applied to the central and outer electrodes, respectively, and, in addition, AC voltages $V_{\rm AC}$ are applied to the outer electrodes as described in the main text.
The outer electrode lines broaden out away from the center in the contact region [Fig.~S1(a)]. The insulating layer [Fig.~S1(b)] is made by depositing 200~nm insulating SiO$_{2}$, leaving a $3 \times 3$~$\mu$m hole over each of the outer electrode line and over the central electrode [Fig.~S1(b)]. The layer of connecting electrodes is made by depositing 200~nm thick indium tin oxide (ITO) on the top [Fig.~S1(c)]. The connecting electrodes provide separate contacts to the central electrode and each of the outer electrodes through the openings in the insulating layer. This allows applying DC voltage to the central and outer electrodes and applying AC voltage to the periodically connected set of outer electrodes (the connection period is seven outer electrodes)~[Fig. S1(d)]. 700~nm wire-bondable Au pads with an ITO underlayer are connected to both the ITO electrodes and the etched area of the sample, creating connections to the carousel electrodes and the ground plane.

\begin{figure}[h]
\centering
\includegraphics[width=3in]{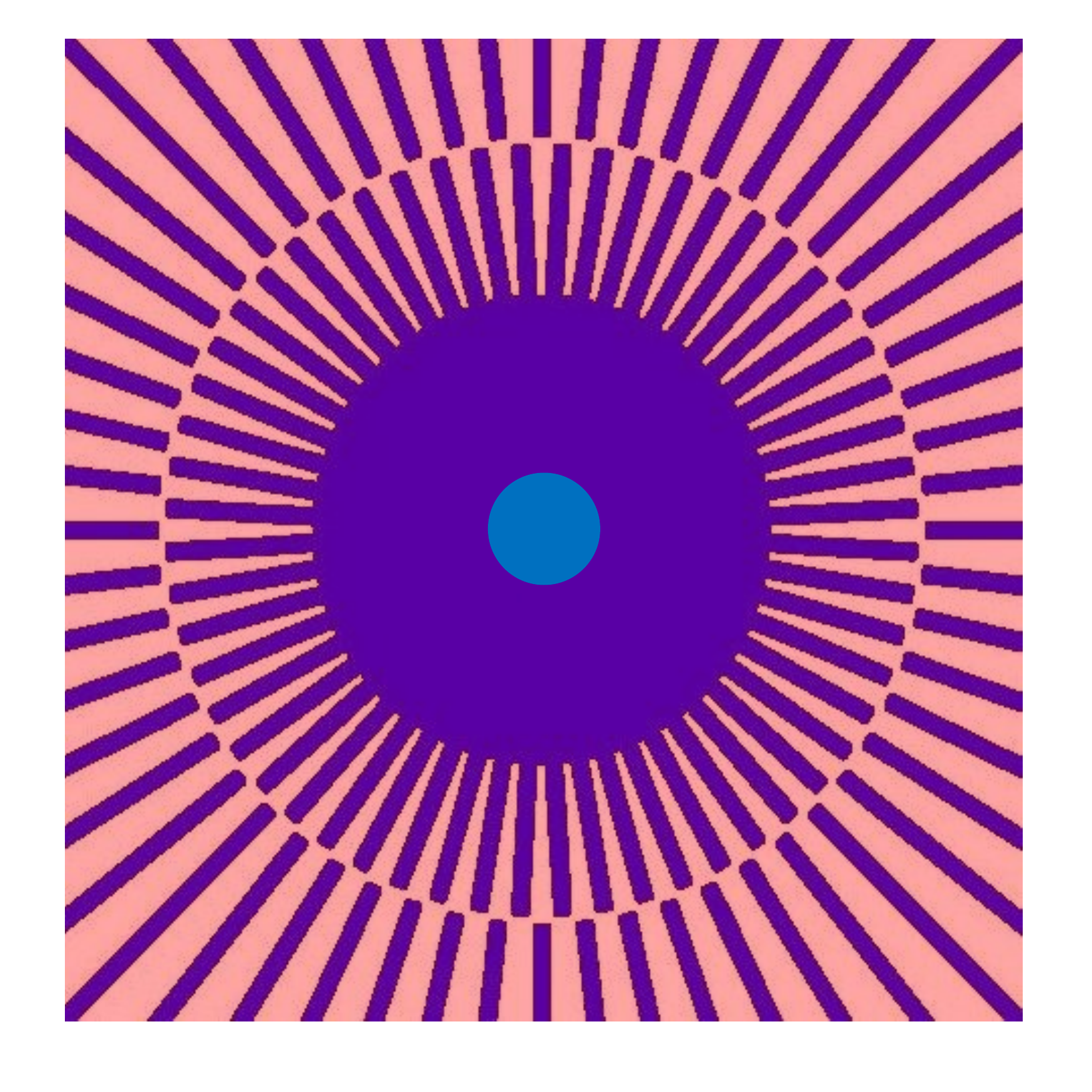}
\caption{Magnified view of the central part of the carousel electrodes and insulating layer. A hole in the center of the insulating layer is for the contact to the central electrode.
}
\end{figure}

\subsection{Circuit schematic of the carousel electronic system}

Figure S3 shows the circuit diagram for the electronic system used to deliver the AC and DC voltages to the carousel electrodes at cryogenic temperatures. AC voltages to the carousel electrodes on the sample are delivered via seven $\sim 1$~m-long broadband transmission lines with impedance-matching termination at the sample. We used coaxial cables UT-141B-SS with silver-plated beryllium copper inner conductor, PTFE Teflon dielectric, and stainless-steel outer shell with diameter 3.6 mm, having a room-temperature attenuation of 3~dB/m at 10~GHz. The cable bandwidth complies with the frequency used in the experiments, while the cable composition reduces heat conductance to the sample. The transmission lines are capacitively terminated to block DC heating at the termination resistors. The DC biases to the central and outer electrodes are supplied separately via regular wires.
%MATT CHECK THIS: A sample resistance between the top and bottom exceeds 10~M$\Omega$ in the experiments.

\begin{figure}[h]
\centering
\includegraphics[width=7in]{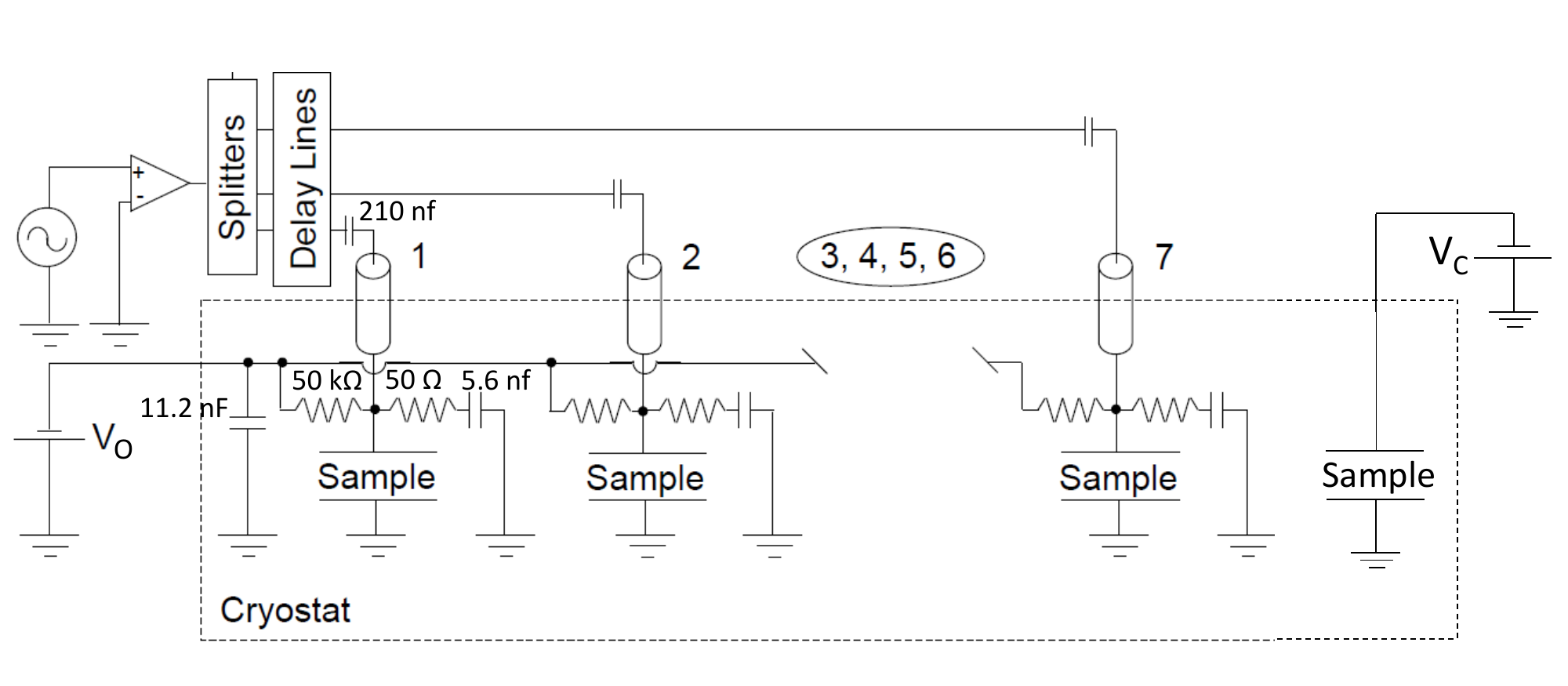}
\caption{Circuit schematic for the carousel. Seven electrode lines apply AC voltages to seven sets of the outer carousel electrodes. In addition, DC voltages $V_{\rm c}$ and $V_{\rm o}$ are applied to the central and outer electrodes, respectively.}
\end{figure}